\documentclass{article}
\usepackage{spconf,amsmath,graphicx,amssymb,ctable}
\usepackage{enumitem}
\usepackage{makecell}
\setlist{nosep, leftmargin=14pt}

\usepackage{mwe} 

\title{Deep Ultrasound Denoising Without Clean Data}
\name{Sobhan Goudarzi*, Hassan Rivaz\thanks{*Corresponding author,
		E-mail address: sobhan.goudarzi@concordia.ca}}
\address{Department of Electrical and Computer Engineering, Concordia University, Montreal, QC, Canada.}
\begin{document}
%
\maketitle
\begin{abstract}
On one hand, the transmitted ultrasound beam gets attenuated as propagates through the tissue. On the other hand, the received Radio-Frequency (RF) data contains an additive Gaussian noise which is brought about by the acquisition card and the sensor noise. These two factors lead to a decreasing Signal to Noise Ratio (SNR) in the RF data with depth, effectively rendering deep regions of B-Mode images highly unreliable. There are three common approaches to mitigate this problem. First, increasing the power of transmitted beam which is limited by safety threshold. Averaging consecutive frames is the second option which not only reduces the framerate but also is not applicable for moving targets. And third, reducing the transmission frequency, which deteriorates spatial resolution. Many deep denoising techniques have been developed, but they often require clean data for training the model, which is usually only available in simulated images. Herein, a deep noise reduction approach is proposed which does not need clean training target. The model is constructed between noisy input-output pairs, and the training process interestingly converges to the clean image that is the average of noisy pairs. Experimental results on real phantom as well as \textit{ex vivo} data confirm the efficacy of the proposed method for noise cancellation. 
\end{abstract}
\section{Introduction}
\label{sec:sec1}
Medical ultrasound image formation pipeline includes two main steps. First, a set of piezoelectric crystal elements of the probe are fired using designed excitation pulses to transmit a specific beam shape into the medium. Second, the backscattered echoes from scatterers are collected, digitized, and beamformed to construct the final image. Depending on the imaging technique (i.e., focused, plane-wave, or synthetic aperture), the transmission and receive steps might be repeated several times to form a single image.\par
\begin{figure}[t!]
	\centering
	\centerline{\includegraphics[width=8.5cm]{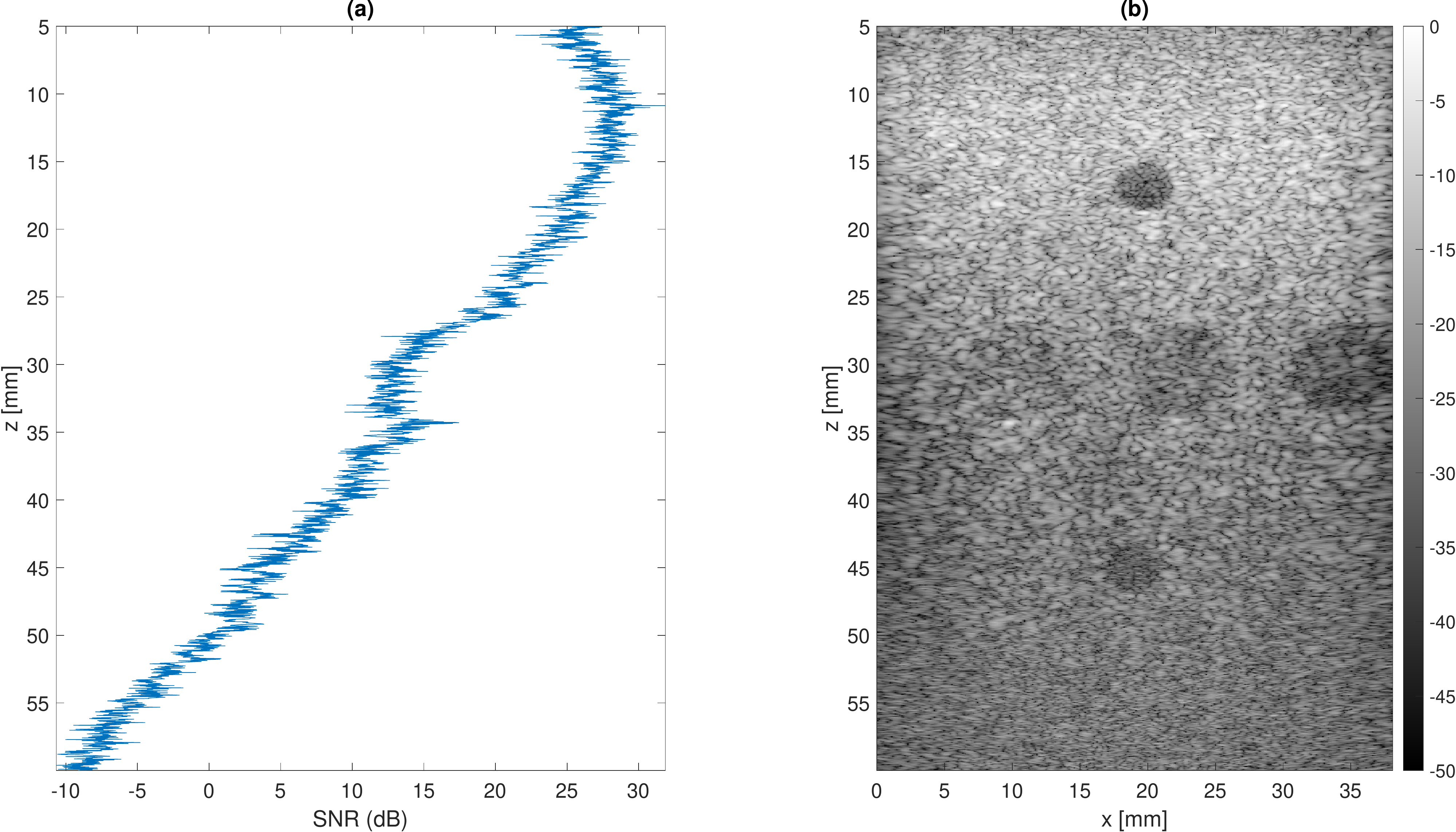}}
	\caption{Experimental measurement of SNR versus depth. (a) Average SNR of RF lines of each depth in dB. (b) B-Mode image of the phantom.}
	\label{fig:fig1}
\end{figure}

The received ultrasound Radio-Frequency (RF) data contains an additive Gaussian noise which is brought about by the acquisition card and the sensor noise~\cite{4982678,wagner1987}. Hence, the quality of resulting image directly relates to the Signal-to-Noise Ratio (SNR) of echo traces acquired by transducer elements. As the transmitted ultrasound beam moves away from the probe surface, it gets attenuated with depth and, consequently, received backscattered signals from deep regions have lower SNR as compared to shallow ones~(Fig.~\ref{fig:fig1}). Therefore, noise reduction of RF data is necessary to have an acceptable quality in deep regions of ultrasound images.\par
By increasing the transmission power, the SNR level can be kept high enough even in the deep regions. However, this approach is limited to the safety threshold preventing damaging increase in thermal index of tissue. The other approach commonly used to remove the noise term, of deep regions, is to average consecutive frame of the same location. This technique not only comes at the expense of reduction in the framerate but also causes blurring artifacts because of unavoidable hand tremors or biological motions. Therefore, lower excitation frequencies are commonly used for imaging deep structures, which reduce the image resolution. Herein, a novel approach based on deep learning for denoising of ultrasound RF data is proposed.\par
Deep learning has widely been used to map noisy observations to the desired clean images in many applications~\cite{MOHDSAGHEER2020102036,TIAN2020251}. The common approach is to train a deep model in a supervised manner using noisy-clean pairs. Clean training targets, however, are not always available since collecting high SNR ultrasound RF data requires either a high power exposure or averaging consecutive frames which is not practical for imaging moving organs such as the heart. Therefore, simulations or phantom experiments are often used to obtain the training data, which create the domain shift problem when performing inference on real data\par
Lehtinen \textit{et al.}~\cite{lehtinen2018} proposed an approach for noise cancellation called Noise2Noise in which clean training target is not required. Their principal finding is that corrupting the training target with zero-mean noise does not change what the network learns. More specifically, for a variety of noise distributions, the clean image is simply the mean, median, or mode of so many noisy observations. Therefore, by selecting appropriate loss function and training with a high enough number of noisy input-output pairs drawn from corruption distribution, deep models can recover the clean image~\cite{lehtinen2018}. This idea has been further extended in subsequent papers~\cite{Krull_2019_CVPR,pmlr-v97-batson19a}.\par
In this paper, the Noise2Noise idea is utilized to remove the additive noise of ultrasound RF data. To this end, a deep model is trained with noisy input-output pairs. As the noise distribution is Gaussian~\cite{4982678,wagner1987}, the Mean Square Error (MSE) loss function is used so that the network learns to output the average of corrupted training data. The experiments are performed on real phantom as well as \textit{ex vivo} data. Results confirm that the proposed approach substantially improves the poor SNR of backscattered signals from deep regions. The efficiency of proposed method is also demonstrated while there are hand tremors among consecutive frame of the same location. 
  
\section{Methods}
\label{sec:sec2}
\subsection{Attenuation model}
\label{sec:sec21}
The power of transmitted ultrasound beam is a function of the depth. Attenuation is the main cause of loss of energy in an ultrasound wave propagating in the tissue, and originates from several mechanisms including wavefront divergence, absorption, and scattering~\cite{cobbold2006}. It can be very complicated to mathematically model how exactly those parameters interact in an inhomogeneous tissue. Nevertheless, a simple mathematical model has been proposed to explain the amplitude ($A$) of the ultrasound beam traveling from probe surface ($z=0$) to depth $z$ as following~\cite{9284544}:
\begin{equation} 
\label{eq:1}
A(f;z,x) = exp(-4\int_{0}^{z}\alpha(f;z^{'},x)dz^{'}) ,
\end{equation}
where $x$ and $f$ denote the lateral position and the frequency, respectively. And $\alpha$ is the attenuation coefficient.\par
As can be understood from Eq.~\ref{eq:1}, the attenuation increases with depth and the beam amplitude goes down. Considering a constant acquisition card and sensor noise levels, the SNR of digitized backscattered signals decreases with depth. This point is experimentally verified on RF data collected from a multi-purpose multi-tissue Ultrasound phantom (CIRS model 040GSE, Norfolk, VA) using Verasonics Vantage 256 platform and the linear L11-5v probe (Verasonic Inc., Kirkland, WA, USA).
The result is illustrated in Fig.~\ref{fig:fig1} and, as can be seen, the average SNR of RF lines falls off with depth. 
\subsection{Noise2Noise training}
\label{sec:sec22}
Let us assume that we have a set of $n$ noisy ultrasound images $\{r_1, r_2, ..., r_n\}$ collected from a fixed medium $v$. The common approach for estimating the clean unknown image $v$ is to find $\widehat{v}$ that minimizes the expectation ($\mathbb{E}$) of deviation from the observations based on a specific loss function ($L$):
\begin{equation} 
\label{eq:2}
\arg\min_{ \widehat{v}} \mathbb{E}_{r}\{L(r,\widehat{v})\} ,
\end{equation}
It can be easily proven that for common $L_2$ and $L_1$ loss functions, the maximum likelihood solutions are the arithmetic mean and median of the measurements, respectively.\par 
Training a deep neural network as a regression model on a set of training input-output pairs $(r_i,v_i)$ is a generalization of above point estimation method:
\begin{equation} 
\label{eq:3}
\arg\min_{\theta} \mathbb{E}_{(r,v)}\{L(f_{\theta}(r),v)\} ,
\end{equation}
where $f$ denotes the network parametrized by $\theta$. For $L_2$ loss function, Eq.~\ref{eq:3} can be rewritten as following:
\begin{equation} 
\label{eq:4}
\arg\min_{\theta} \frac{1}{n} \sum_{i=1}^n(f_{\theta}(r_i)-v_i)^2 ,
\end{equation}
It has been discussed in~\cite{lehtinen2018} that when the mapping between input-output pairs is not 1:1 (e.g., in super resolution problem), the network learns an average of all plausible functions through minimization of Eq.~\ref{eq:4}. This tendency has been skillfully employed for training deep denoising models when the clean target is not available~\cite{lehtinen2018}. More specifically, it has been shown that a network trained for mapping noisy images to noisy images converges to output the average image. Therefore, for a variety of noise distributions, if we select an appropriate loss function, the clean image can be recovered by only looking at noisy observations.
\subsection{Proposed denoising model}
\label{sec:sec23}
As shown in Fig.~\ref{fig:fig1}, the noise level of ultrasound data increases with depth which makes the bottom part of B-Mode images useless. Herein, we take benefits of explained method in Section~\ref{sec:sec22} for denoising the ultrasound RF data. More specifically, if we collect $n$ frames from a medium, each frame can be considered as a noisy observation of the medium. Consequently, there are $C(n,2)$ pairs of noisy images from the same medium. By collecting multiple frames from different mediums, a deep regressor can be trained on resulting noisy input-output pairs. And, as will be shown in the Results Section, the network learns to output each noisy input to an average of all noisy observations which is the desired clean image of medium.\par
Herein, the U-Net architecture~\cite{ronneberger2015u} is used to construct a deep regressor between noisy input-output pairs. To keep the generalization performance, prevent overfitting of the denoising model and reduce computational and memory loads, the number of kernels in each convolutional layer of original U-Net architecture is divided by a factor of 4. As a result, a light U-Net with only 1.08 million parameters (16 times smaller) has obtained. As the noise distribution of ultrasound RF data is Gaussian~\cite{wagner1987}, the $L_2$ loss function is used to force the model to output the mean of all noisy observations. PyTorch library is used for implementing the model. The batch size is selected as 8, and AdamW optimizer with $\beta_1 = 0.9$ and $\beta_2 = 0.999$ is used. The network is trained for 100 epochs. Although the low SNR is not an issue in shallow regions of ultrasound images, the network is trained on whole image to prevent blocking artifact. Fortunately, learned mapping function does not have any negative effect on high SNR regions and mainly affects the low SNR parts of the image.       
\section{Experiments}
\label{sec:sec3}
\subsection{Dataset}
\label{sec:sec31}
\begin{figure}[t!]
	\centering
	\centerline{\includegraphics[width=6cm]{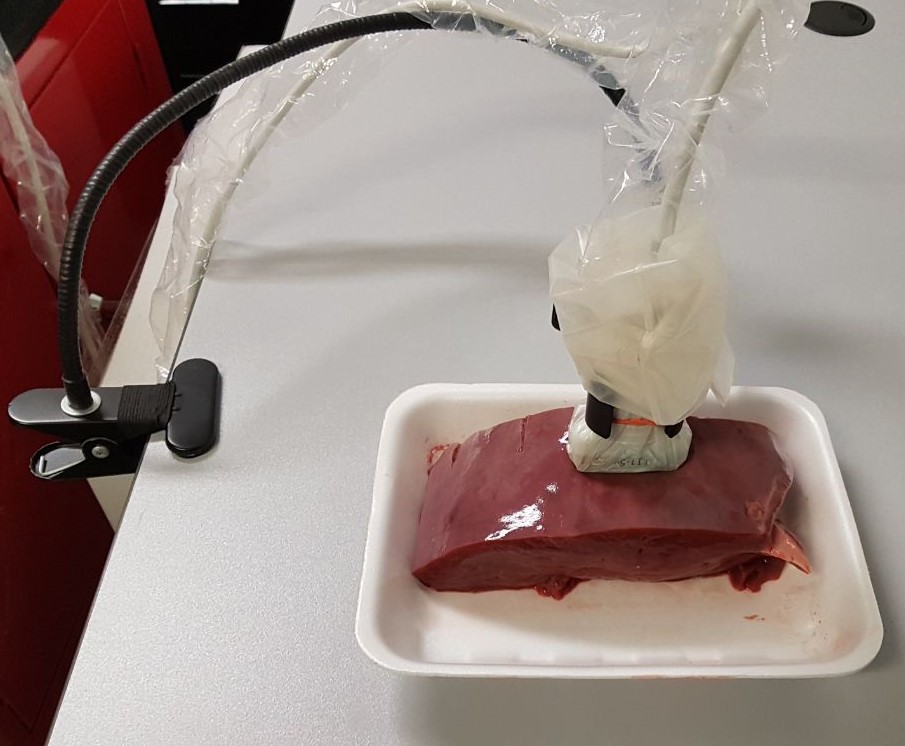}}
	\caption{The real \textit{ex vivo} experiment setup used for collecting ultrasound images from bovine liver.}
	\label{fig:fig2}
\end{figure}
The whole datasets have been collected using Verasonics Vantage 256 platform and the linear L11-5v probe (Verasonic Inc., Kirkland, WA, USA). Ultrasound images were collected using the plane-wave technique with a single $0^o$ transmission. The center frequency of probe was set to 7.6 MHz, and RF data have been digitized with the sampling frequency of 31.25 MHz. The images were reconstructed using the Delay-and-Sum (DAS) beamforming.\par
Two sets of phantom and \textit{ex vivo} data were collected. The phantom was a multi-purpose multi-tissue Ultrasound phantom (CIRS model 040GSE, Norfolk, VA), and bovine liver as well as chicken breast have been used for \textit{ex vivo} experiments. For each image, 30 consecutive frames were collected while the probe was fixed using a mechanical arm to prevent any movement. Each frame is a noisy observation of the field.
Fig.~\ref{fig:fig2} shows the real \textit{ex vivo} experiment setup. General purpose ultrasound probe cover is used for cleaning purposes.\par
\subsection{Evaluation settings}
\label{sec:sec32}
To quantitatively assess the performance of the proposed method, common indexes including Structural Similarity (SSIM), Normalized Root Mean Square Error (NRMSE), and Peak Signal to Noise Ratio (PSNR) are reported. As a ground-truth image is required for index calculation, the average of 100 frames, collected without any relative motions of target in connection with probe, is considered as clean image.
Training has only been completed using phantom data and \textit{ex vivo} data kept for test step. In total, the training dataset includes 1350 noisy input-output pairs from which $90\%$ was used for training and $10\%$ for validation. The network weights of each epoch are saved during training. Once the training step is completed, the network with best validation error is selected for test step. 
\section{Results}
\label{sec:sec4}
\begin{figure}[t!]
	\centering
	\centerline{\includegraphics[width=8cm]{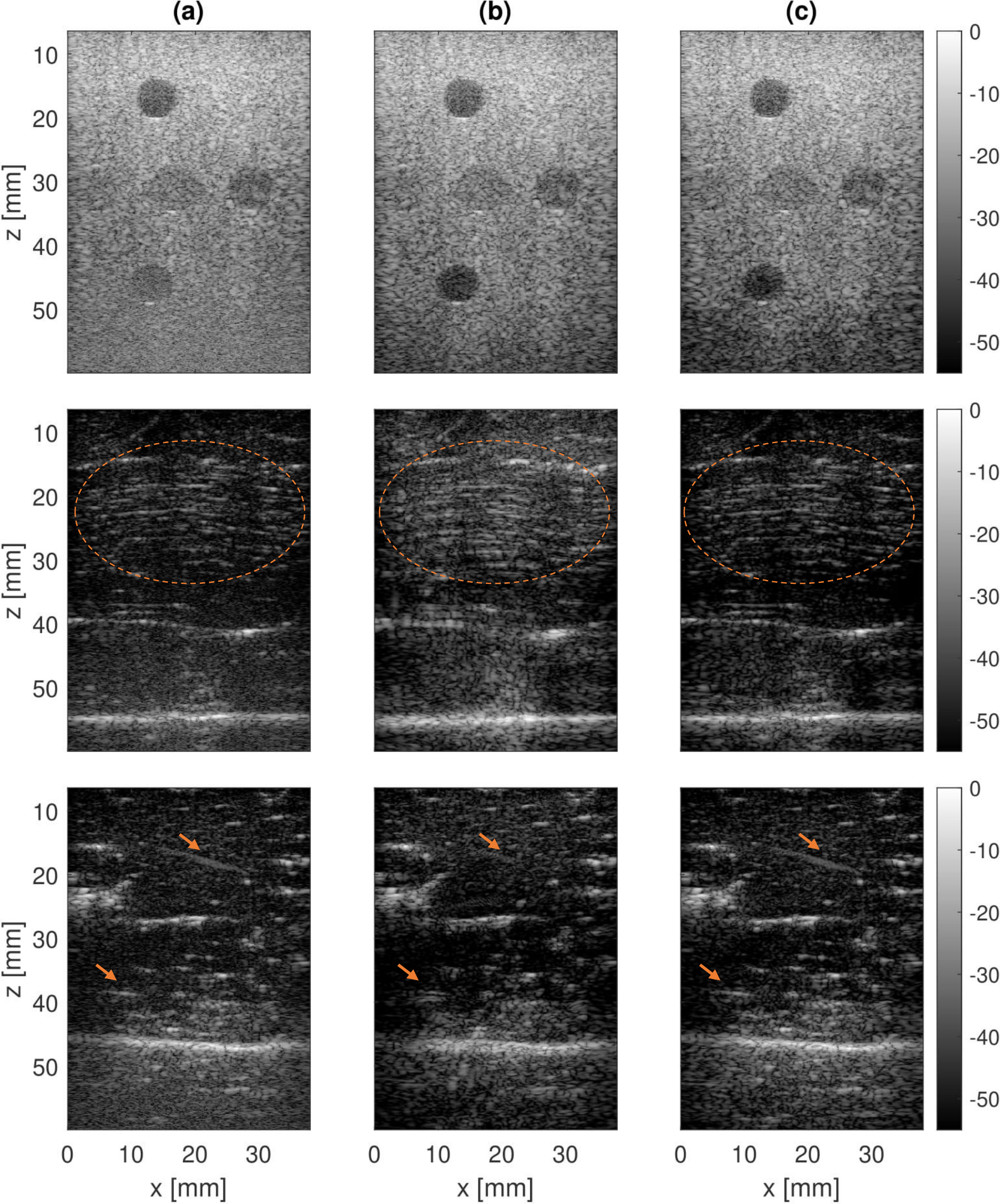}}
	\caption{The results of the proposed method on test datasets. (a) B-Mode image of a single frame. (b) Average B-Mode of 30 frames. (c) The output of the proposed method. First row indicates the phantom data while second and third rows correspond to chicken breast and bovine liver data, correspondingly.}
	\label{fig:fig3}
\end{figure}

\begin{table}[b!]
	\caption{Quantitative results in terms of PSNR, NRMSE, and SSIM indexes for experimental test data. Best performances are marked with bold font.}
	\label{table:1}
	\centering
	\setlength{\tabcolsep}{2.5pt}
	\scriptsize
	\begin{tabular}{c c c c c c c c c c c c c c c c c}
		\specialrule{.15em}{0em}{.2em}
		dataset & phantom & chicken breast & bovine liver   \\ [.2em] 
		\specialrule{.05em}{0em}{.2em} 
		index & PSNR NRMSE SSIM & PSNR NRMSE SSIM & PSNR NRMSE SSIM \\ [.2em] 
		\specialrule{.05em}{0em}{.2em} 
		\makecell{noisy input \\ averaging frames \\ proposed method} & \makecell{25.11 \\ 36.37 \\ \textbf{37.27}} \makecell{0.024 \\ 0.007 \\ \textbf{0.006}} \makecell{0.76 \\ 0.87 \\ \textbf{0.9}} & \makecell{29.75 \\ 32.39 \\ \textbf{37.11}}  \makecell{0.016 \\ 0.012 \\ \textbf{0.008}} \makecell{0.85 \\ 0.86 \\ \textbf{0.91}}& \makecell{27.73 \\ 29.72 \\ \textbf{31.28}} \makecell{0.02 \\ 0.018 \\ \textbf{0.017}}  \makecell{0.82 \\ 0.84 \\ \textbf{0.88} }\\ [.2em] 
		\specialrule{.05em}{0em}{.2em}
	\end{tabular}
\end{table}
To evaluate the generalization performance of trained model, it is blindly tested on unseen phantom and \textit{ex vivo} data without any further fine-tuning. The achieved improvement is demonstrated in comparison with the noisy input image. Moreover, the result of the proposed method is compared with the averaging approach in which 30 consecutive frames are averaged to remove the noise effect. Fig.~\ref{fig:fig3} illustrates the results.\par 
As can be observed in the first row of Fig.~\ref{fig:fig3}, the proposed method successfully removes the noise of the input ultrasound image. The averaging approach is also able to completely cancel the noise term of phantom ultrasound image because all frames are perfectly aligned. Besides the fact that averaging approach reduces the framerate, it is also not practically applicable while there are unavoidable motions of the target with respect to the probe, such as hand tremor or biological motions. In order to prove this point, a small pressure has been applied while collecting \textit{ex vivo} data. Therefore, successive frames are not perfectly aligned. The second and third rows of Fig.~\ref{fig:fig3} demonstrate that the misalignment causes blurring artifacts after averaging, while our method is not affected since its input only contains a single frame of data. The averaging artifact on chicken breast and bovine liver images is highlighted using orange dash line and arrows, respectively.\par
The result of quantitative comparison is presented in Table~\ref{table:1}. All indexes confirm the effectiveness of proposed method in removing noise. The result of averaging approach is comparable to our results on phantom data because of not having any misalignment. For \textit{ex vivo} results, however, our method works far better than averaging.  
\section{Conclusions}
\label{sec:sec5}
The attenuation of transmitted ultrasound beam with depth and the additive Gaussian noise of collected RF data are the main reasons associated with low quality of ultrasound images in deep regions. Herein, the proposed deep model removes the noise term by only looking at noisy observations. The experimental results show that the training converges to clean image of medium and does not suffer from misalignment between consecutive frames. This approach is especially useful when clean training target is not available.  
\section{Acknowledgments}
\label{sec:sec6}
This work was supported by Natural Sciences and Engineering Research Council of Canada (NSERC) RGPIN-2020-04612.
%

\bibliographystyle{IEEEbib}
\bibliography{strings,refs}
\end{document}